\documentclass[a4paper,11pt]{article}
\usepackage{pos}

\title{Measurements of lepton-jet azimuthal decorrelation and 1-jetttiness event shape at high $Q^2$ in deep inelastic scattering (DIS) with the H1 experiment at HERA}

\author*[a]{Sookhyun Lee}

\affiliation[a]{Physics Department, University of Michigan,\\
450 Church St., Ann Arbor, USA}

\emailAdd{sookhyun@umich.edu}

\abstract{

Recent progress towards realizing the Electron-Ion Collider (EIC)~\cite{AbdulKhalek:2021gbh} that was announced in 2020 to be built at Brookhaven National Laboratory (BNL) in the United States has revived interests in the deep-inelastic-scattering (DIS) data taken from $ep$ collisions at Hadron-Electron Ring Accelerator (HERA). The H1 detector equipped with tracking and fully instrumented calorimeter detectors in its hermetic and asymmetric design is well suited for jet and event shape measurements. Proposed measurements involving jets, global events shapes and correlations in DIS for HERA and EIC focus on the three-dimensional description of nucleon structure and hadronization and their their flavor dependence as well as precision measurements for QCD and BSM. This contribution presents recent results on the lepton-jet azimuthal decorrelation and the 1-jettiness event shape measurements performed using the DIS data at high $Q^2$ taken with the H1 detector at HERA.  }

\FullConference{%
  *** Particles and Nuclei International Conference - PANIC2021 ***\\
  *** 5 - 10 September, 2021 ***\\
  *** Online ***
}

\begin{document}
\maketitle
%

\section{Introduction}
HERA was the first $ep$ collider located in Hamburg Germany that brought electrons (or positrons) with a beam energy of 27.6~GeV and protons with a beam energy of 920 GeV to collisions at a corresponding centre-of-mass energy of 319~GeV. 
Notably, the analysis of DIS data collected at HERA pioneered ways in which we understand the proton structure, e.g. in terms of the dynamics of its constituents, parton distribution functions (PDFs). Various structure functions measured at HERA were used to extract PDFs with great precision \cite{herapdf}. 
On the other hand, a new direction has been taken such that a full three dimensional description of the constituent partons inside the proton, adding the transverse momentum dependent (TMD) component~\cite{sivers} to the original picture in which partonic motions inside the proton were collinear and/or taking into account effects of higher twist such as multi-parton correlations~\cite{stermanqiu} in order to explain the results of various spin-momentum correlation measurements in hadronic collisions have become the main focus in hadronic physics.
In the new picture, the precise determination of the strong coupling constant $\alpha_s$, a fundamental parameter in QCD, at different scales using the DIS data is critical in achieving an overall improved accuracy in predictions of various particle production processes at hadronic colliders.\\
\indent During the data taking period from 1994 to 2007, the H1 detector~\cite{h1detector} recorded approximately 200 $pb^{-1}$ in $e^{-}p$ collisions and $\sim$ 300 $pb^{-1}$ in $e^+p$ collisions. The main detector components for the measurements discussed here are the Central Jet Chambers (CJC), a tracking system providing a $p_T$ resolution less than 1\%, and the LAr and Spacal Calorimeters covering a wide pseudorapadity range between -3.82 and 3.35 with an energy resolution less than 4\% for both electrons and hadrons.  

\section{Lepton-Jet Azimuthal Decorrelations}

With the theoretical advancement achieving an accuracy in pQCD computation at next-to-next-leading-order (NNLO) and beyond for jet productions in $pp$ as well as $ep$ collisions, jets have become an increasingly attractive observable as a tool to probe the intricate structure of the nucleon and dynamics of hadronization process. In DIS, the lepton-jet azimuthal decorrelation defined as the magnitude of the momentum imbalance vector $\lvert \vec{q}_{T} \rvert = \lvert \vec{k}^{e}_{T} + \vec{P}^{jet}_{T}\rvert$, where $\vec{k}^{e}_{T}$ and $\vec{P}^{jet}_{T}$ are the momentum of a scattered electron and a jet projected onto the plane perpendicular to the beam axis, respectively, has been proposed as a noble and clean way to access the intrinsic transverse motion of the quarks inside the proton~\cite{Liu:2018trl}. 
Defined in this manner, the theory calculation does not involve non-perturbative (NP) TMD fragmentation functions (FFs) that describe the hadronization process inside the jets. It is sensitive to TMD quark PDFs $f_q$ via the relation implied in the factorization formula shown in Eq.~\ref{eq:ljet}, where $H_{TMD}$ is the hard factor and $S_J$ is the soft factor. 
\begin{equation}
\frac{d^5\sigma (lp\rightarrow l'p)}{dy^e d^2k^e_T d^2q_T} = \sigma_0 \int d^2k_{T}d^2\lambda_{T} xf_q(x,k_T,\mu_F) H_{TMD}(Q,\mu_F) S_J \delta^{(2)}(q_T-k_T-\lambda_T)
\label{eq:ljet}
\end{equation}
The unfolding of data for detector effects were performed on four variables, the jet $p_T$ and $\eta$, the azimuthal decorrelation $\Delta \phi$ and $q_T$, simultaneously using a machine learning based method Omnifold~\cite{omnifold}.
\subsection{Results}
\begin{figure}[hbt!]
\centering
\includegraphics[width=.46\textwidth]{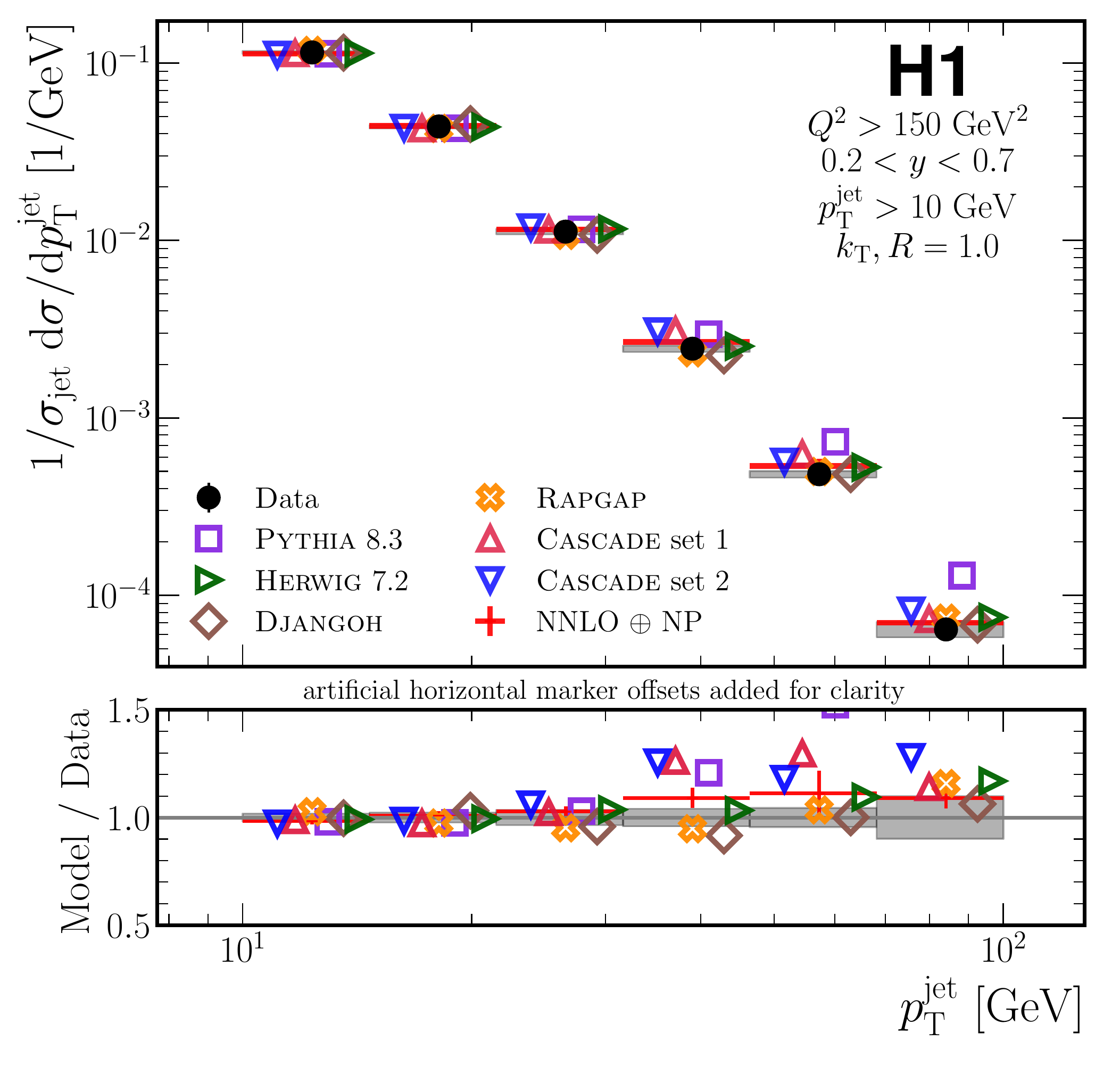}
\includegraphics[width=.46\textwidth]{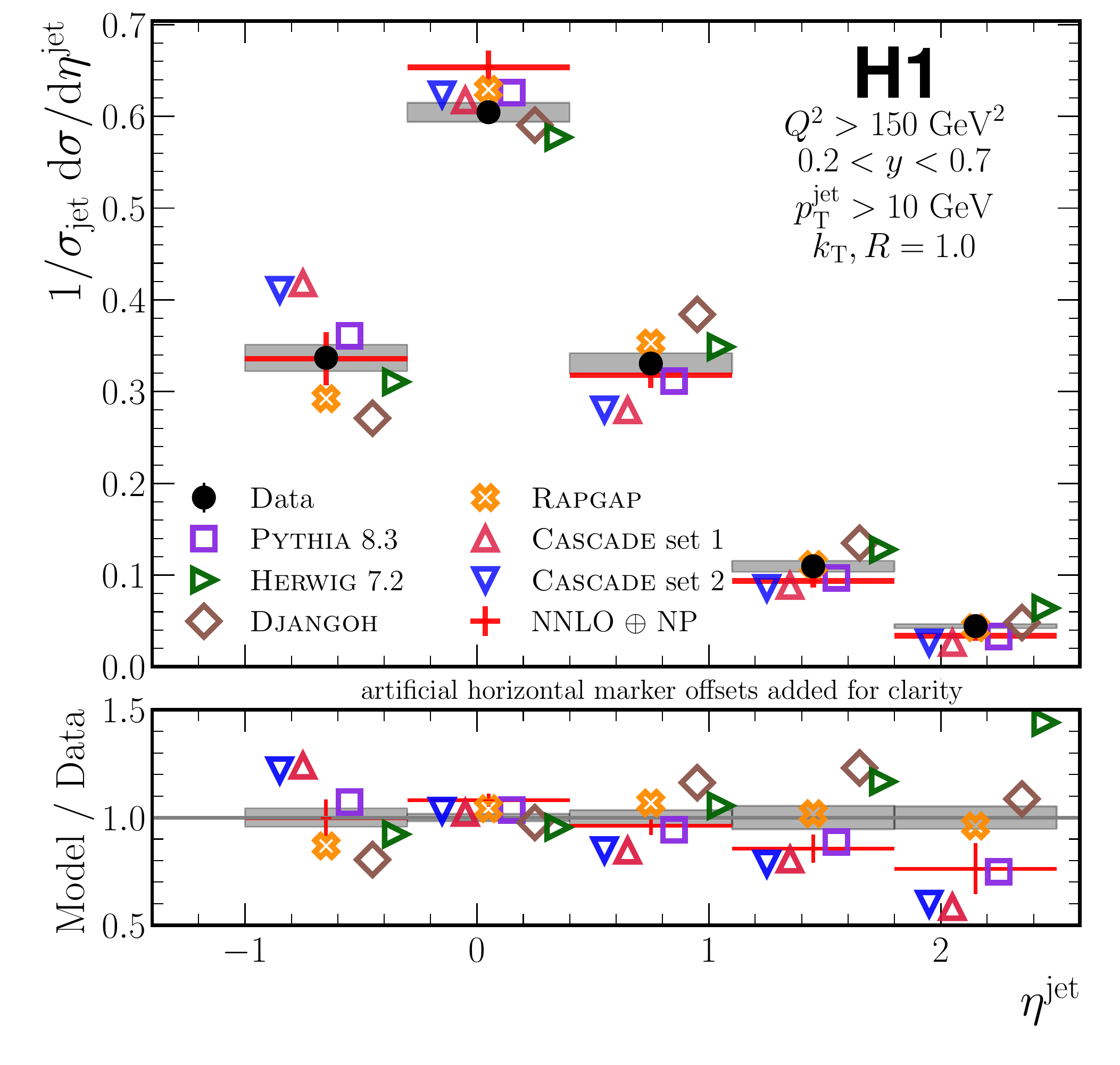}
\caption{ The jet $p_T$ and jet $\eta$ distributions from~\cite{ljetcorr}.}
\label{fig:jetpteta} 
\end{figure}

The measured transverse momentum ($p_T$) of jets in Fig.~\ref{fig:jetpteta} is reasonably well described by NNLO pQCD calculation corrected for NP effects and a slight deviation is seen with increasing pseudorapidity ($\eta$). Rapgap~\cite{Rapgap} (using LO matrix element with LL parton shower) of all predictions describes the data best. An overall shape difference is seen for Cascade (TMD based)~\cite{Cascade} in both variables.    

\begin{figure}[h!]
\centering
\includegraphics[width=.46\textwidth]{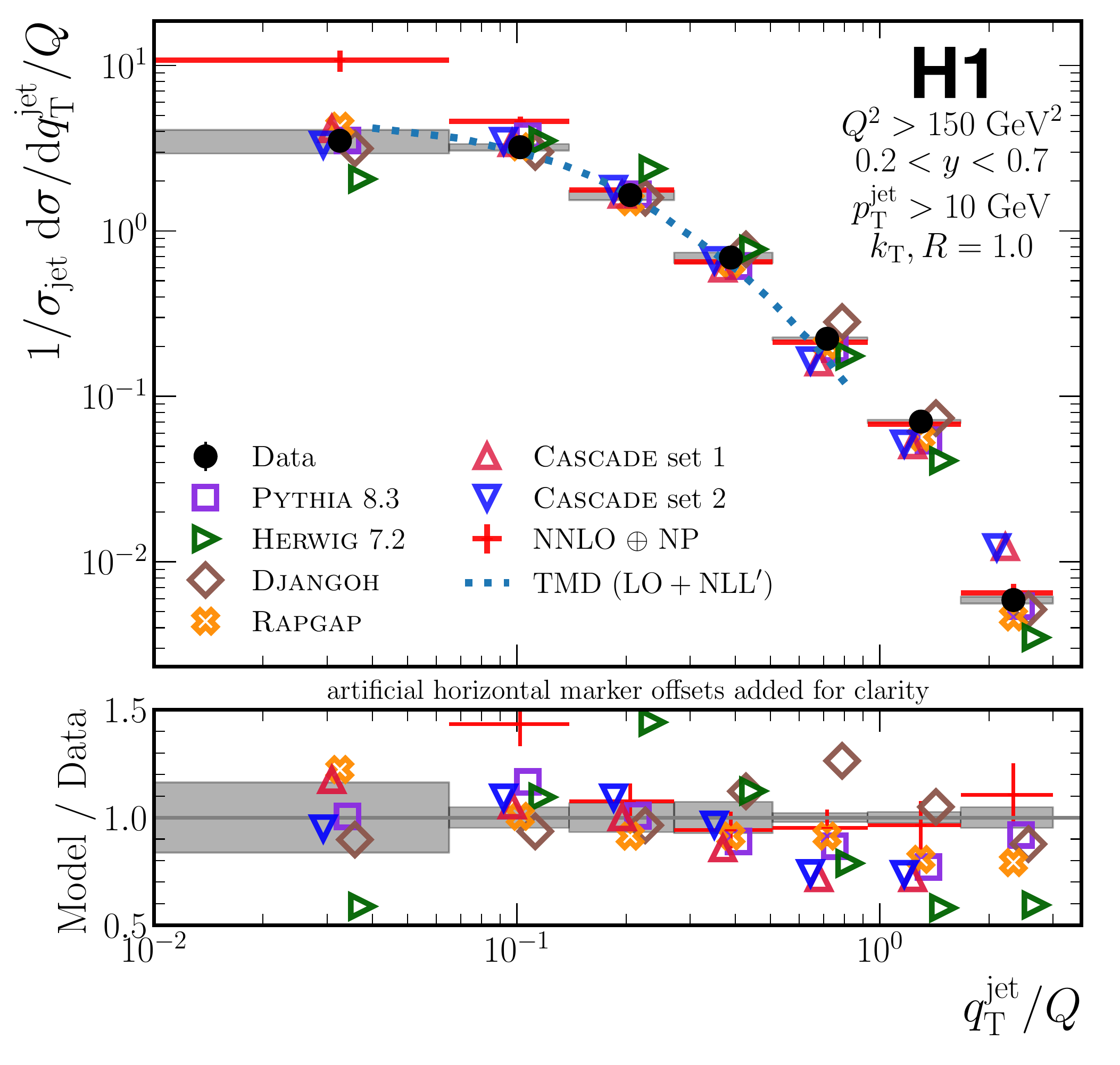}
\includegraphics[width=.46\textwidth]{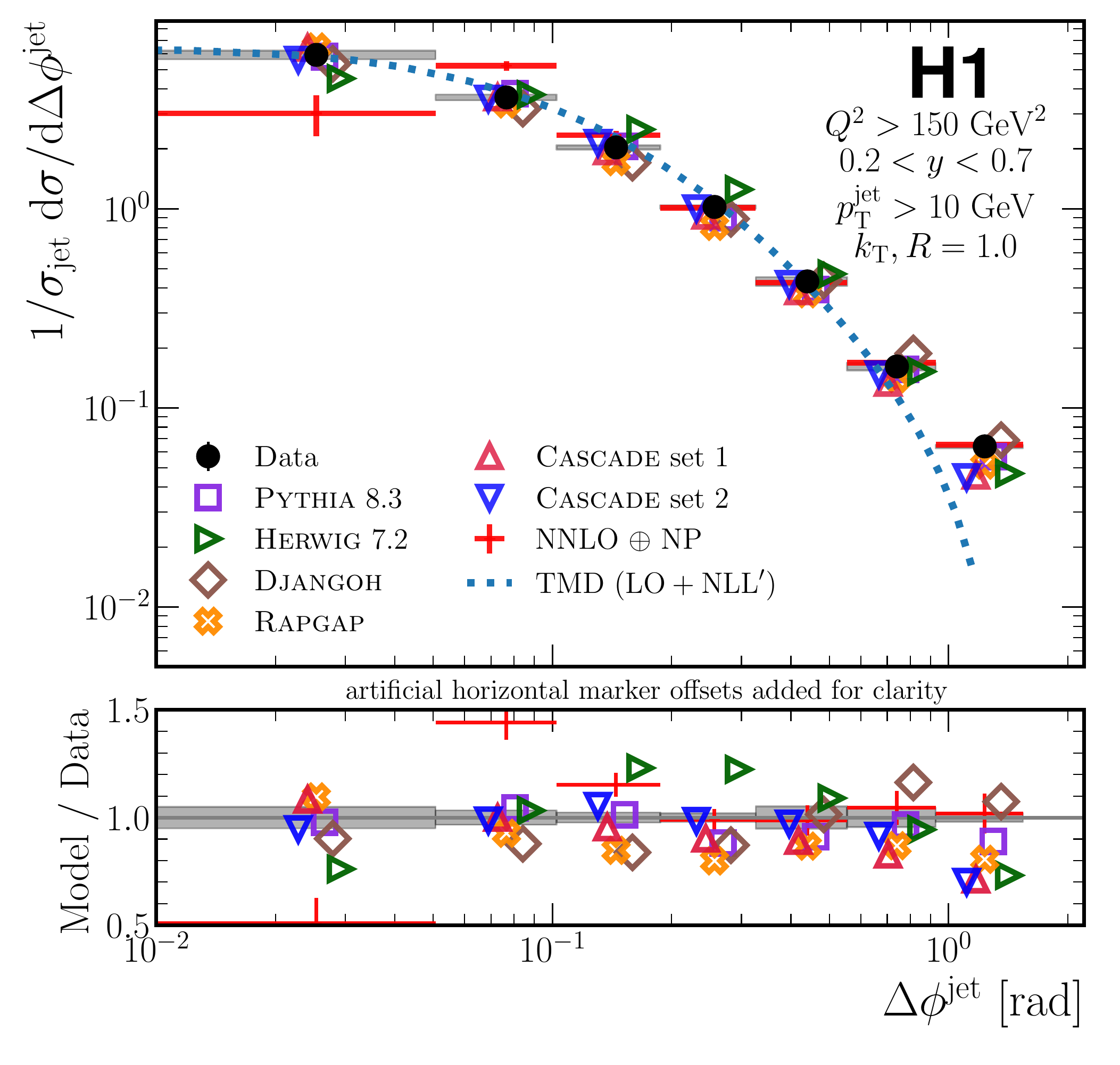}
\caption{ The $q_T$ and $\Delta \phi$ distributions from~\cite{ljetcorr}.}
\label{fig:qt} 
\end{figure}

The measured $q_T$ in Fig.~\ref{fig:qt} is reasonably well described by NNLO at higher $q_T$, by both NNLO and TMD between 0.2 and 0.6 (well beyond the validity region of $>$ 0.25 for TMD) and only by TMD in the lowest region. A similar pattern is shown for $\Delta \phi$. A large overlap covered by data will help constrain the matching between the TMD and higher twist collinear pQCD frameworks.

\section{1-jettiness event shape}
An event shape observable 1-jettiness, denoted as $\tau_{1}^b$, proposed to be measured at HERA as well as EIC~\cite{Kang:2013nha} has been measured using the H1 data at HERA. The 1-jettiness $\tau_{1}^b$ defined in Eq.~\ref{eq:tau1b} is a dimensionless, Lorentz invariant, global and infrared-safe observable. It is expected to provide great constraining power on the strong coupling constant $\alpha_s$ and NP hadronization effects.
\begin{equation}
\tau_1^b = \frac{2}{Q^2} \Sigma_{i \in X } \quad min \{ (xP+q) \cdot p_i, xP \cdot p_i \}
\label{eq:tau1b}
\end{equation}
The denominator $Q$ denotes the square-root of the virtuality $Q = \sqrt{-q^2}$, where $q$ is the momentum of an exchanged boson. The index $i$ runs over all of the particles in an event and $p_i$ is the four-momentum of each particle. The first term in the bracket projects $p_i$ onto the jet axis that is defined as the sum of the scattered quark momentum $xP$ and $q$, the second term onto the beam axis $xP$. 

\subsection{Results}
The $\tau_{1}^b$ distribution in inclusive DIS features an interesting structure where there is a distinctive broad peak at small $\tau_1^b$ and a tail with an additional delta-function like peak at one. The $\tau_1^b \rightarrow 0$ limit corresponds to events with two pencil-like jets, while high $\tau_1^b$ region is dominated by multi-jet events. At $\tau_1^b = 1$, the current hemisphere is empty. 

\begin{figure}[tbh!]
\centering
\includegraphics[width=.47\textwidth]{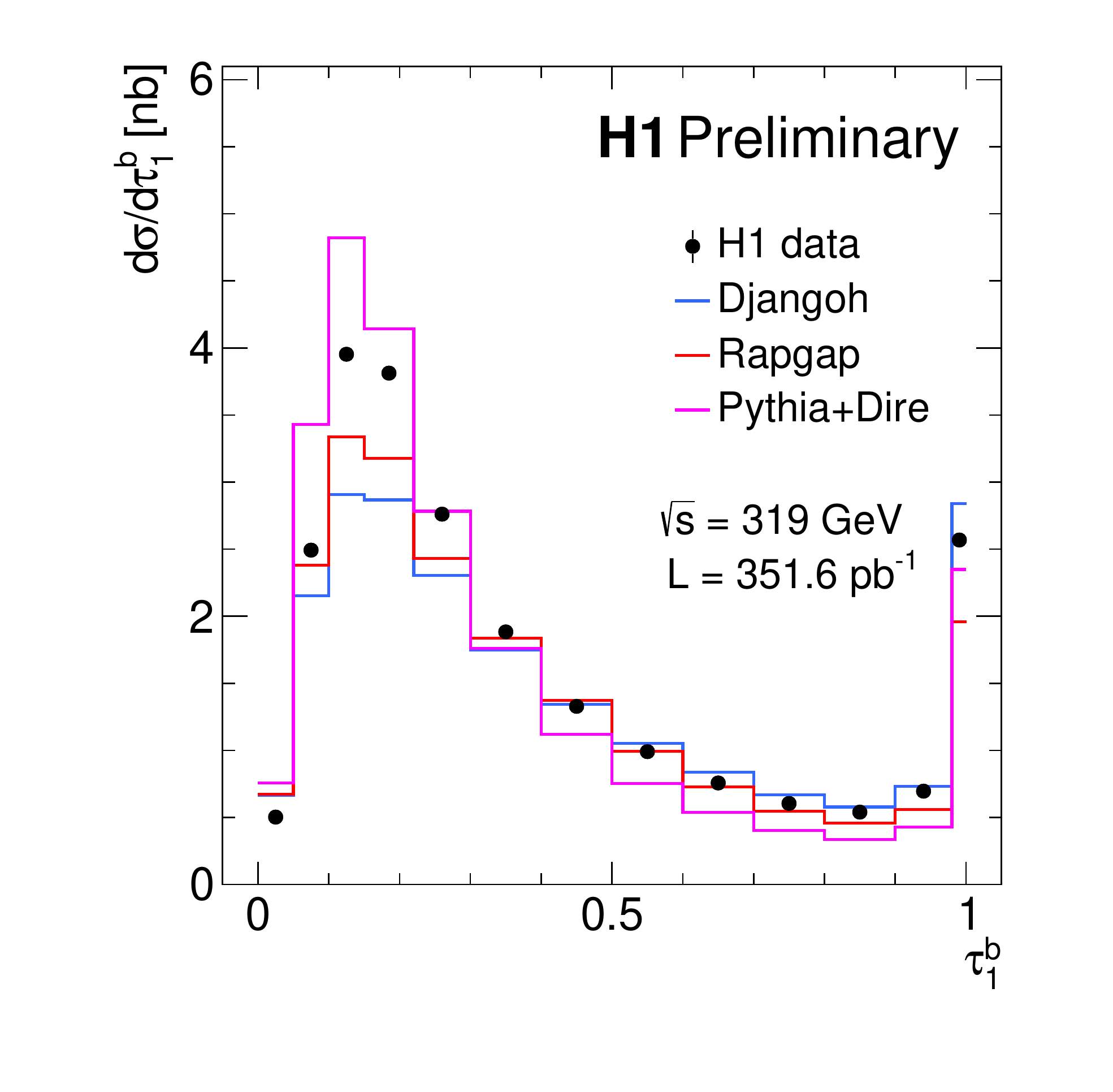}
\includegraphics[width=.47\textwidth]{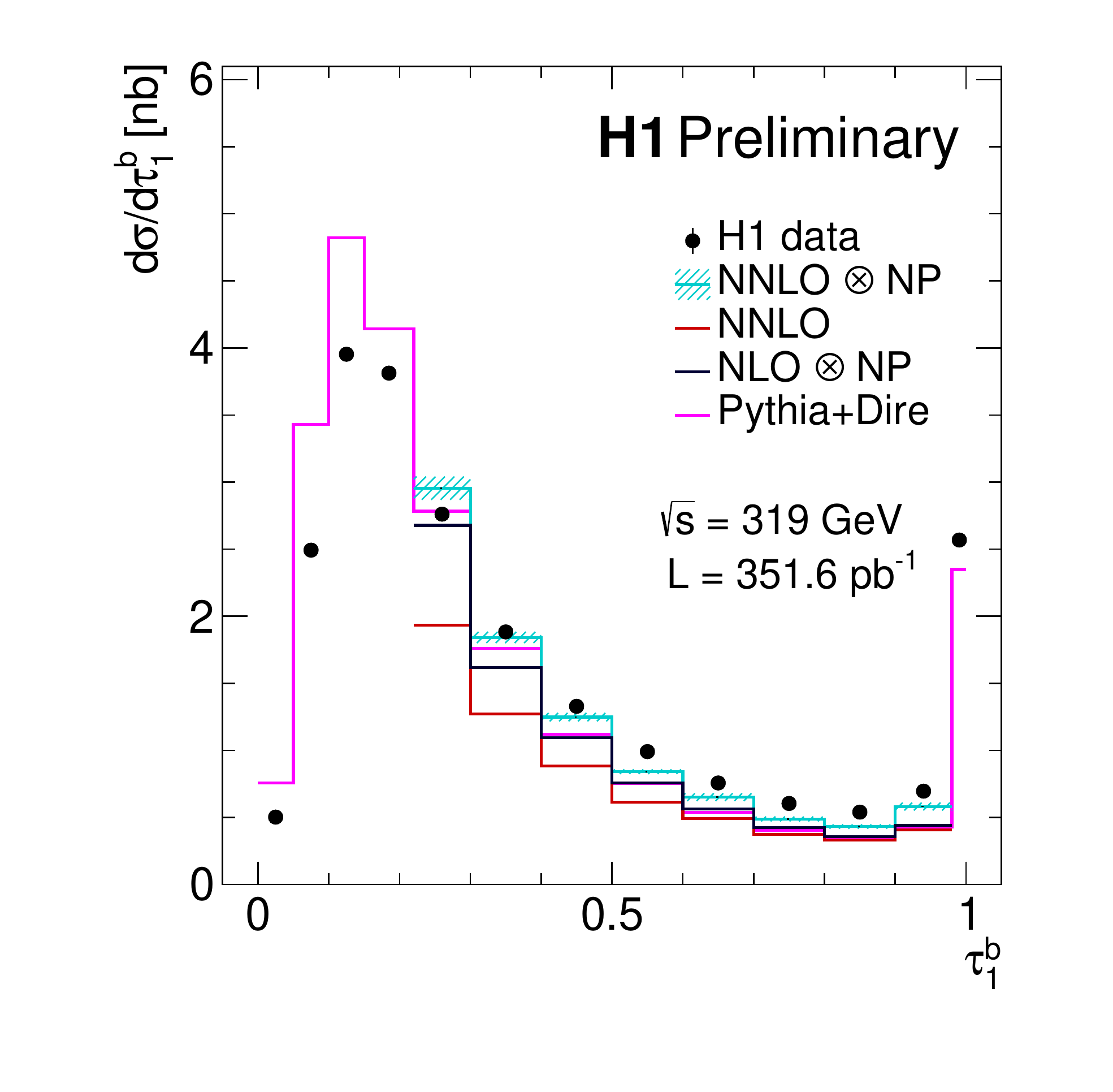}
\caption{ The $\tau_1^b$ distribution in inclusive DIS data compared with (left) MC simulations and (right) NLO (NNLO) pQCD computations with/witout NP corrections.}
\label{fig:tau1bincl} 
\end{figure}

In Fig.~\ref{fig:tau1bincl}, the measured data shows an excellent agreement with Django~\cite{Django} and Rapgap~\cite{Rapgap} in mid-to-high $\tau_1^b$ region. An improved agreement with data is seen at higher order (NNLO) in high $\tau_1^b$ region where the fixed-order (FO) pQCD calculations predominantly determine the shape of the distribution. In contrast, none of the simulations describe the $\tau_1^b$ in the peak region where NP effects are large. In the tail region, including NP effects in pQCD calculation better describes the data. Pythia8~\cite{Pythia} with Dire parton shower shows poor agreement.

\begin{figure}[tbh!]
\centering
\includegraphics[width=.495\textwidth]{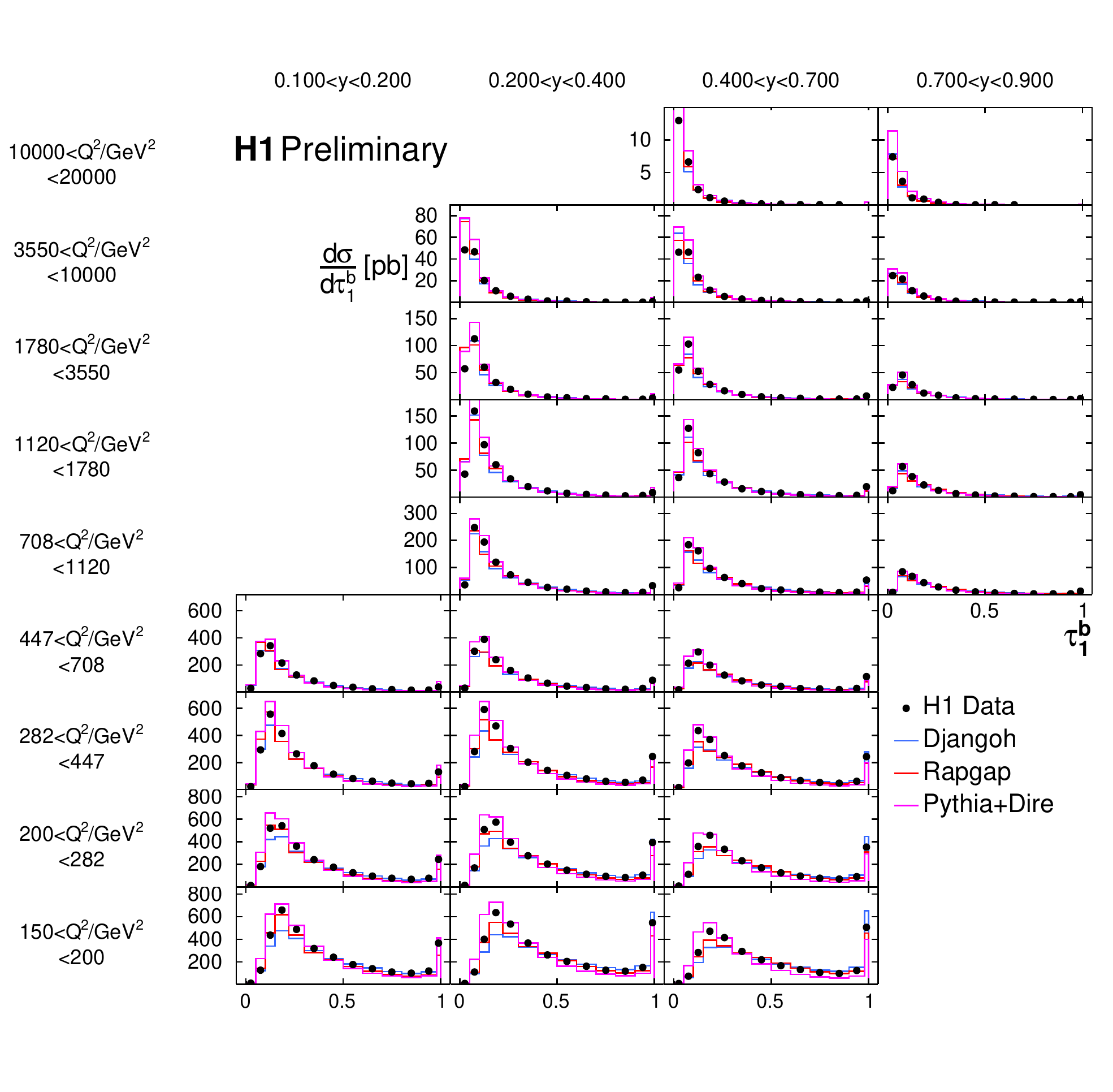}
\includegraphics[width=.495\textwidth]{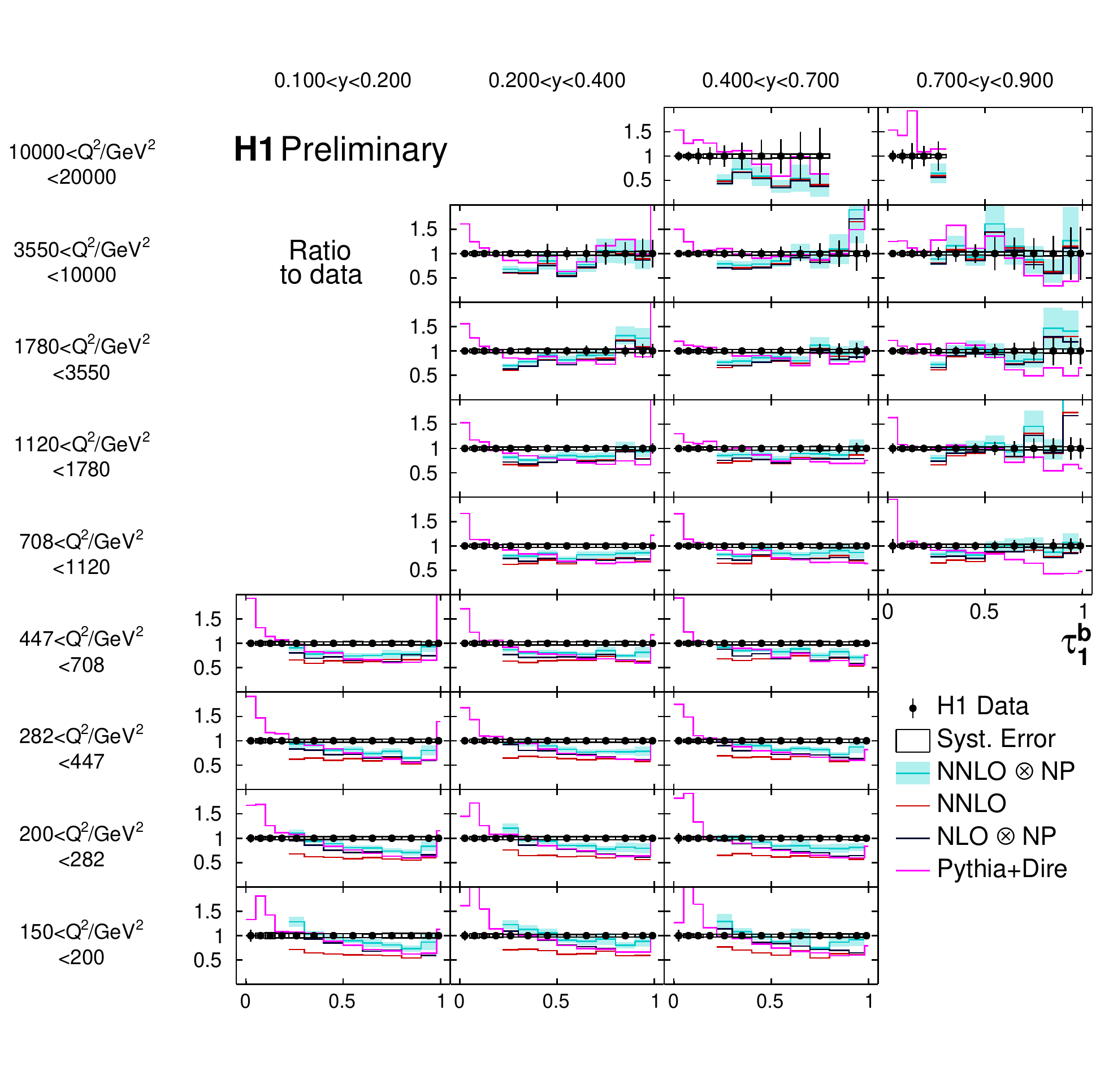}
\caption{The $Q^2$ and $y$ dependent $\tau_1^b$ distributions (left) and comparisons with pQCD calculations (right).}
\label{fig:tau1btriple} 
\end{figure}

The shape of the $\tau_1^b$ depends strongly on $Q^2$ and $y$. In Fig.~\ref{fig:tau1btriple}, on top is shown the $\tau_1^b$ distributions for different $Q^2$ and inelasticity $y$ bins. Jets tend to be increasingly collimated with $Q^2$ and the shift of the peak position caused by NP effects is more prominent at low $Q^2$. In high $Q^2$ and high $y$ region where higher order and NP effects are suppressed, the pQCD prediction describes the data best as shown on the right in Fig~\ref{fig:tau1btriple}. 


\section{Summary and outlook}
First measurements of lepton-jet azimuthal decorrelation and 1-jettiness have been made using the H1 DIS data taken at HERA. The novel method of using jets to access partonic transverse motions inside the proton proved to be sensitive at the center-of-mass energy of HERA to the kinematic region where a reconciliation of the two theoretical approaches can be tested in the transition from the non-perturbative TMD regime into the higher twist collinear pQCD regime. A global event shape $\tau_1^b$ measured at various $Q^2$ and $y$ values accentuates effects of dominant physics process in the region, e.g. fixed-order, parton shower and non-perturbative hadronization. When measured in inclusive DIS, it provides sensitivity to PDFs. Predictions at N$^3$LL resummation accuracy in soft-collinear effective theory (SCET) framework will become available soon. A combined analysis of NNLO and N$^3$LL computation and fully triple differential cross sections measurements with respect to $y$, $Q^2$ and $\tau_1^b$ will constrain with unprecedented precision $\alpha_s$ that governs the accuracy of theoretical predictions at different scales as well as non-perturbative physical quantities extracted from data. Follow-up measurements are under way and will serve as a path-finder towards the EIC.


\end{document}